\documentclass[sn-mathphys-num]{sn-jnl}

\usepackage[T1]{fontenc}
\usepackage[utf8]{inputenc}
\usepackage{textcomp}
\usepackage{graphicx}%
\usepackage{multirow}%
\usepackage{amsmath,amssymb,amsfonts}%
\usepackage{amsthm}%
\usepackage{mathrsfs}%
\usepackage[title]{appendix}%
\usepackage{xcolor}%
\usepackage{textcomp}%
\usepackage{manyfoot}%
\usepackage{booktabs}%
\usepackage{algorithm}%
\usepackage{algorithmicx}%
\usepackage{algpseudocode}%
\usepackage{listings}%

\theoremstyle{thmstyleone}%
\theoremstyle{thmstyletwo}%

\theoremstyle{thmstylethree}%

\lstdefinelanguage{json}{
    basicstyle=\ttfamily\tiny,
    numbers=left,
    numberstyle=\tiny\color{gray},
    stepnumber=1,
    numbersep=8pt,
    showstringspaces=false,
    breaklines=true,
    frame=single,
    backgroundcolor=\color{lightgray!20},
    literate=
     *{0}{{{\color{blue}0}}}{1}
      {1}{{{\color{blue}1}}}{1}
      {2}{{{\color{blue}2}}}{1}
      {3}{{{\color{blue}3}}}{1}
      {4}{{{\color{blue}4}}}{1}
      {5}{{{\color{blue}5}}}{1}
      {6}{{{\color{blue}6}}}{1}
      {7}{{{\color{blue}7}}}{1}
      {8}{{{\color{blue}8}}}{1}
      {9}{{{\color{blue}9}}}{1}
      {:}{{{\color{red}:}}}{1}
      {,}{{{\color{red},}}}{1}
      {\{}{{{\color{magenta}\{}}}{1}
      {\}}{{{\color{magenta}\}}}}{1}
      {[}{{{\color{magenta}[}}}{1}
      {]}{{{\color{magenta}]}}}{1},
}

\usepackage[labelfont=bf]{caption}
\begin{document}

\title[Article Title]{Non-Fungible Blockchain Tokens for Traceable Online-Quality Assurance of Milled Workpieces}


\author*[1]{\fnm{Nicolai} \sur{Maisch}}\email{nicolai.maisch@isw.uni-stuttgart.de}

\author[1]{\fnm{Shengjian} \sur{Chen}}

\author[1]{\fnm{Alexander} \sur{Robertus}}

\author[1]{\fnm{Samed} \sur{Ajdinović}}

\author[1]{\fnm{Armin} \sur{Lechler}}

\author[1]{\fnm{Alexander} \sur{Verl}}

\author[1]{\fnm{Oliver} \sur{Riedel}}

\affil[1]{\orgdiv{University of Stuttgart}, \orgname{ISW}, \orgaddress{\street{Seidenstraße 36}, \city{Stuttgart}, \postcode{70174}, \state{Baden-Württemberg}, \country{Germany}}}


\abstract{This work presents a concept and implementation for the secure storage and transfer of quality-relevant data of milled workpieces from online-quality assurance processes enabled by real-time simulation models. 
It utilises Non-Fungible Tokens (NFT) to securely and interoperably store quality data in the form of an Asset Administration Shell (AAS) on a public Ethereum blockchain. 
Minted by a custom smart contract, the NFTs reference the metadata saved in the Interplanetary File System (IPFS), allowing new data from additional processing steps to be added in a flexible yet secure manner.
The concept enables automated traceability throughout the value chain, minimising the need for time-consuming and costly repetitive manual quality checks.}


\maketitle

\section{Introduction}
The modern manufacturing industry faces significant challenges from volatile markets and dynamic customer demands. 
The increasing need for customised products and flexibility in small-batch production requires highly adaptable processes. 
Adherence to stringent quality standards is crucial, as component quality is the benchmark for the performance and reliability of production systems~\cite{VDI.2015}. 
Traditional quality assurance processes, however, are often resource-intensive and costly. 
Manufacturing companies use comprehensive measurement procedures to ensure compliance with quality specifications, while recipient companies frequently perform their own checks to identify defects. 
This duplication within the supply chain results in high costs and operational inefficiencies~\cite{mitra2016fundamentals}.  
A promising approach to enhancing quality assurance efficiency is usage of real-time simulation combined with an online-quality assurance model. 
By generating a virtual workpiece using real-time data, this method enables continuous online-quality monitoring, automates inspection, and could support extensive mechanical measurements or manual sample-based checks~\cite{chen2024utilizing}.

A key element of quality assurance is traceability across multiple manufacturers, with the objective of targeting corrective actions in response to quality issues. 
Blockchain technology offers a decentralised, tamper-resistant, and transparent solution for traceable data storage.
Initially popularised by cryptocurrencies such as Bitcoin, it is now being adopted in supply chain management systems due to its security and transparency benefits~\cite{Agarwal.2022}. 
To enable secure data exchange and collaborative management while ensuring privacy compliance, tokenisation of real-world assets offers a promising solution. 
Tokens in blockchains are digital assets that represent and enable the secure digitisation of real-world information for transparent ownership or data transfer.
Beyond digital ownership in art and gaming, tokens could carry product data from manufacturing processes. 
This enables automated, secure transfer of quality information, eliminating redundant inspections as recipients can reliably access manufacturing data to assess component quality efficiently.

This study examines the integration of real-time simulation and non-fungible blockchain tokens to enhance automated online-quality assurance in CNC milling processes on the use case of industrial robots. 
It explores the use of blockchain technology and tokenisation as a tamper-proof method for recording quality attributes, and presents an efficient concept for workflow automation. 
Implemented on an industrial robot, the concept automates the creation of digital quality records, enabling traceability in flexible manufacturing systems down to batch size one. 
Validation through milling trials confirms the effectiveness of the framework and demonstrates its potential to streamline quality assurance and improve traceability in modern manufacturing.

\section{State of the Art} \label{ch:sota}

\subsection{Online-quality assurance} \label{ch:quality-assurance}
\begin{figure}[!ht]
    \centering
    \includegraphics[width=0.88\textwidth]{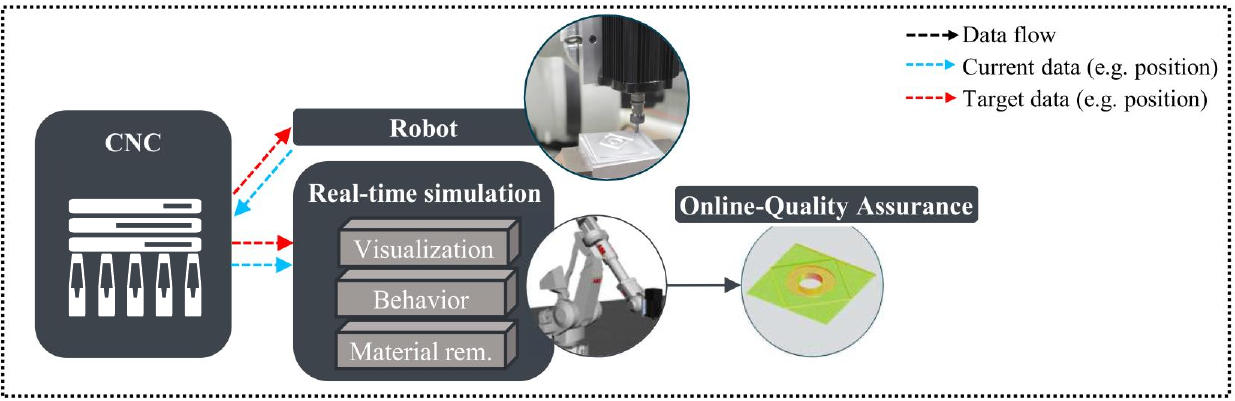}
    \caption{Online-quality assurance framework with an industrial robot use case}
    \label{fig:online-quality}
\end{figure}

Online-quality assurance enables real-time monitoring of the quality of milled workpieces during the manufacturing process. There are different types of online-quality assurance methods, e.g. data based methods~\cite{huang2022edge} or model based methods~\cite{altintas2014virtual}. In this proposed work a method is used by extending real-time simulation models from virtual commissioning to operational-parallel real-time simulation. Thereby is the simulation model directly connected to the actual control, allowing real-time communication~\cite{chen2024utilizing}. 
With data such as actual and target position data, a virtual workpiece can be simulated by using an embedded material removal simulation. 
This virtual workpiece can be seamlessly exported from the real-time simulation to an online-quality assurance service module, as outlined in \cite{chen2024digital}. 
This service module uses a feature-based approach to assess workpiece quality, analysing individual features against shape and positional tolerances defined by DIN 4760:1982-06 and BS 7172:1989. 
Robust detection and segmentation enable standard algorithms to evaluate attributes such as hole roundness or edge straightness, following methods specified in these standards. 
The primary challenge lies in implementing this methodology within an online-quality assurance module, ensuring real-time detection and evaluation~\cite{chen2024utilizing}.
The whole methodology is shown in Fig.~\ref{fig:online-quality}, where the control of an industrial robot is connected to a real-time simulation. The real-time simulation is comprised of three constituent components: namely, visualization model, a behavior model and a material removal simulation. The material removal simulation is responsible for simulating a virtual workpiece, the purpose of which is to facilitate the calculation of quality during the process by means of an online quality assurance service.  
\vspace{-0.2cm}
\vspace{-0.85cm}

\subsection{The AAS for Quality Assurance Data} \label{ch:aas-interopera}
In order to digitally map and successfully exchange quality assurance data between supply chain partners, it is necessary to ensure syntactic and semantic interoperability. 
The Asset Administration Shell (AAS) provides a framework for the standardised mapping of product and machine information \cite{AASiD.2022}. 
The actual data representing the asset is arranged into submodels that address various aspects of the asset's life cycle phase. 
To overcome semantic hurdles and achieve syntactic interoperability via the standardised structure of the AAS, industry standards are being developed to standardise submodel content\footnote{https://industrialdigitaltwin.org/en/content-hub/submodels}. 
This standardised representation of asset information enables seamless exchange between machines and companies.
The submodel \textit{Quality Control for Machining} \cite{InterOperaSubmodel.2024} standardises the information required to represent quality assurance-related information for a milled component, aligning it with existing standards for quality assurance, e.g. ISO 2768:1989-1. 
Therefore, it provides a foundation for cross-company quality assurance of workpieces by storing quality data in a standardised format and exchanging the information interoperably with industry partners.

\subsection{Blockchain Technology for Supply Chain Traceability}
When multiple companies are involved in manufacturing of a workpiece, a traceability framework is essential for quality assurance. 
Blockchain technology enables immutable, decentralised data storage, ensuring transparency and tamper-proof records through cryptographic methods \cite{Nakamoto.2008}. 
It ensures \textbf{authenticity} by tracing data origins, maintains \textbf{integrity} by preventing data changes during transmission and provides \textbf{accountability} by verifying data transmission \cite{Schutte.2017}. 
These features make blockchain a viable solution for trust in data exchange between partners across the supply chain, increasing traceability and transparency by reducing the risk of counterfeiting and tampering.
Furthermore, public blockchains have the potential to enhance efficiency in supply chain management by facilitating access to a unified platform for all stakeholders. \cite{Bodkhe.2020}

Blockchain technologies like \textit{Ethereum} are not just a cryptocurrency but a platform for decentralised applications and smart contracts \cite{Buterin.2014}. 
Smart contracts are programs executed on a blockchain, automating interactions and ensuring trusted, transparent transactions. 
They automatically enforce predefined rules when conditions are met, enabling automation between real-life information systems and the blockchain in supply chain traceability networks.
Another key feature is the ability to create blockchain-based tokens enabling tailored asset management. 
Non-fungible tokens (NFTs) uniquely represent assets or data, with ownership and details stored in metadata \cite{EthereumNft}. NFTs are widely used for managing and exchanging digital artwork\footnote{e.g. https://opensea.io/collection/boredapeyachtclub}.

\subsection{Overview of Blockchain Applications in Traceability Networks}
This section provides a brief overview of blockchain solutions for supply chain traceability, highlighting their advantages and challenges.
Two approaches using customised smart contracts, without relying on a token-based system design, are proposed in \cite{Frey.2019} and \cite{Maisch.2024}. 
The first uses a permissioned blockchain to store material data and reduce redundant measurements, though at the cost of transparency. 
The second enables collaborative work on a public blockchain by storing hashed AAS data, ensuring verifiable yet private supply chain tracking.

The use of NFTs as a manageable, single source of truth for product data has been demonstrated in the context of quality certificates for food specialities \cite{Sung.2022} and the authenticity of medical devices \cite{Gebreab.2022}. 
In the manufacturing domain, \cite{Westerkamp.2020} proposes NFTs to track and trace goods throughout production, focusing on tokenised product composition mechanisms to streamline traceability processes. 
\cite{AlKhader.2023} employs NFTs for efficient supply chain management and traceability, enabling all parties to record product data, manufacturing details and certificates securely. 
In order to incorporate scalability to their NFT solution, \cite{Gebreab.2022, Westerkamp.2020, AlKhader.2023} combine the security features of blockchain technology with the \textit{Interplanetary File System (IPFS)}. 
This enables data to be centrally stored while still being linked in a decentralised way within the blockchain, thus preserving the tamper-proof characteristics \cite{IPFS.2022}.

\section{NFT-based Traceability Platform}
The concepts discussed in Section \ref{ch:sota} demonstrate how blockchain’s transparency and immutability can ensure secure asset traceability in manufacturing, while tokenisation enhances identification and enable seamless transactions through encapsulated data representation. 
However, an automated solution for in-process real-time quality data tokenisation, fully leveraging blockchain’s traceability and data management capabilities for manufacturing, is lacking.
To address this gap, the following section introduces an NFT-based exchange system for online-quality assurance data of milled workpieces. This system enables efficient and secure cross-company data exchange while addressing the key requirements interoperability, traceability, modifiability and privacy.

\begin{figure}[!ht]
    \includegraphics[width=0.66\textwidth]{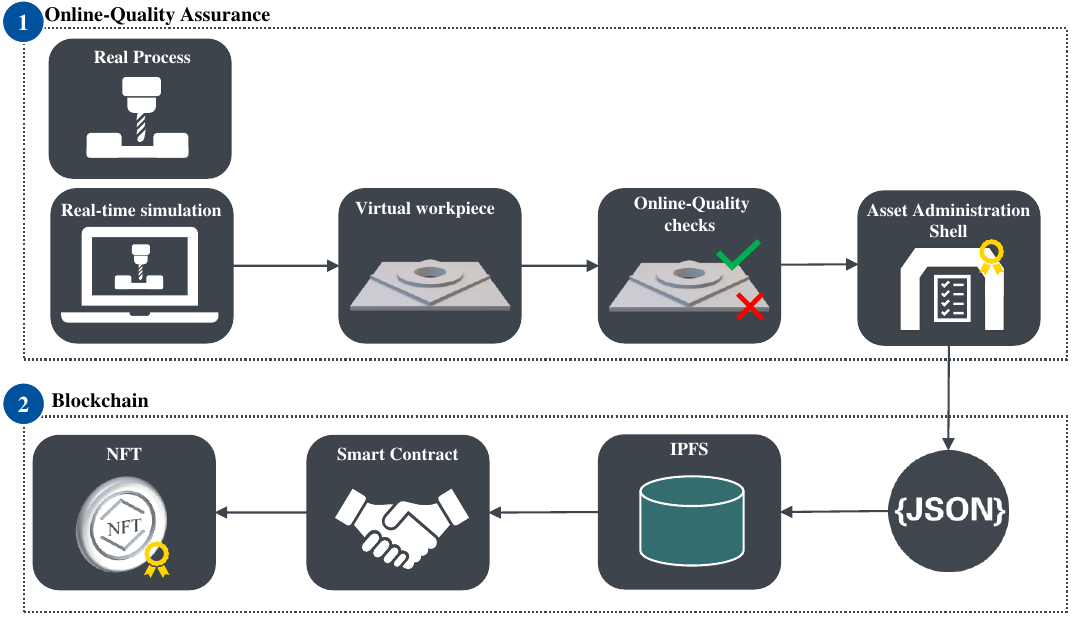}
    \centering
    \caption{Concept overview for secure exchange of quality relevant information}
    \label{fig:concept-overview}
\end{figure}

The online-quality assurance approach (Section \ref{ch:quality-assurance}) results in a virtual workpiece that represents the quality-relevant features by real-time simulation of the milling process. 
In order to exchange this information interoperably across the supply chain, the AAS together with the interopera submodel (Section \ref{ch:aas-interopera}) is used as a standardised information model with standardised semantics (Fig.~\ref{fig:concept-overview}) \cite{InterOperaSubmodel.2024}. 

\vspace{-0.4cm}
\vspace{-0.5cm}
The following blockchain layer enables cross-company management and traceability of quality data.
Within the proposed conceptual framework, Ethereum-based NFTs have been identified as the optimal representation of quality-relevant data, ensuring transparency while leveraging blockchain’s security. 

\begin{figure}[!ht]
    \includegraphics[width=0.62\textwidth]{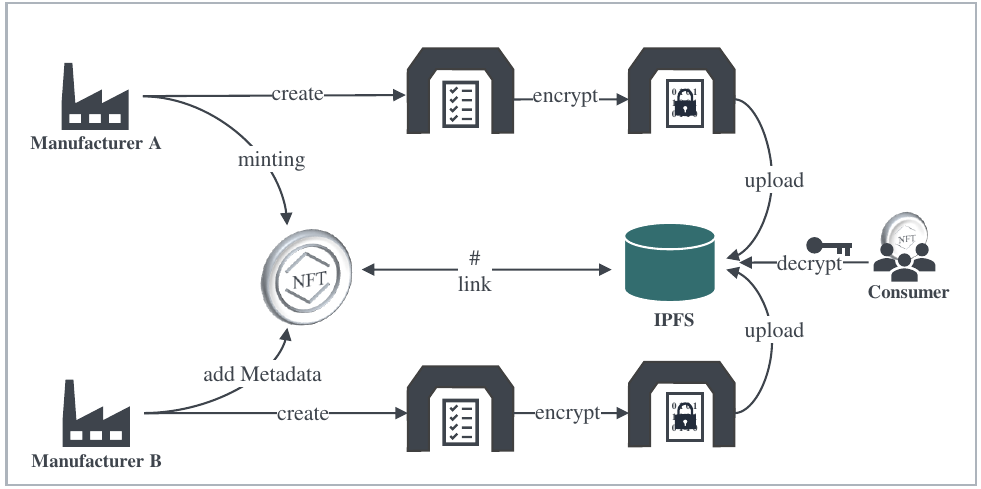}
    \centering
    \caption{Collaborative enrichment of quality-relevant data by the NFT and IPFS}
    \label{fig:colab}
\end{figure}

To address Ethereum’s scalability limitations, the framework integrates IPFS for data storage, with the NFT acting as a pointer by storing the corresponding hash value.
The JSON-serialised AAS from the quality assurance process is uploaded to an IPFS node, making it publicly accessible and traceable through the NFT’s hash. 
During the minting process (i.e. the creation of the NFT), this hash is embedded in the NFT, linking it directly to the stored data. 
If modifications occur - whether through further manufacturing steps or updates to the AAS - the NFT’s hash can be revised accordingly, ensuring continuous traceability~(Fig.~\ref{fig:colab}). 
This approach enables a modifiable, collaborative process while maintaining a reliable audit trail.

Since data on IPFS is publicly accessible, the AAS is encrypted using a symmetric key~(Fig.~\ref{fig:colab}). 
This key is shared only with authorised partners, ensuring data privacy while preserving the benefits of decentralised and immutable storage.

\section{Implementation and Results}
\begin{figure}[!ht]
    \includegraphics[width=0.38\linewidth]{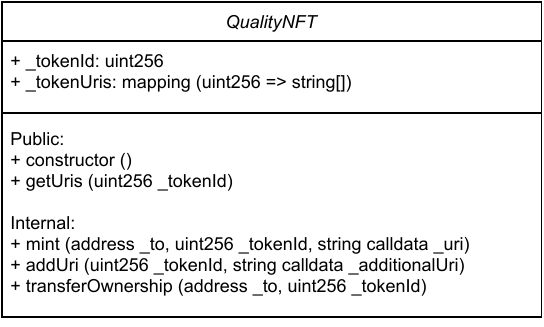}
    \centering
    \caption{UML class diagram of the smart contract for quality assurance NFTs}
    \label{fig:smartContractUml}
\end{figure}

To create a NFT as proof of quality, a smart contract based on the Ethereum standard ERC-721 was developed by using open source libraries\footnote{https://github.com/OpenZeppelin/openzeppelin-contracts}.
The contract incorporates a user-defined token ID, such as a sequential number or the serial number of the product, and maps this token ID to the corresponding IPFS URI stored within the token (Fig.~\ref{fig:smartContractUml}).

\vspace{-0.3cm}
\vspace{-0.5cm}

After deploying the smart contract via its constructor, tokens are minted, specifying an ownership address, the token's ID, and a linked IPFS URI. 
The NFT owner is permitted to add additional IPFS URIs following further manufacturing steps. 
Each added URI points to a JSON file containing part manufacturing quality data. 
In practice, this leads to different tokens within a collection, each representing the quality characteristics of a workpiece by representing the different states at different stages of production.   
If the workpiece itself changes ownership, the NFT can also be transferred from the current owner, giving the new owner permissions for internal functions. 

\vspace{-0.1cm}
\begin{table}[h!]
\centering
\resizebox{\textwidth}{!}{%
\begin{tabular}{cccc}
\hline
\textbf{Feature Name} & \textbf{Actual Size} & \textbf{Deviation from Target Dimension} & \textbf{Within Tolerance} \\ \hline
Height Area 1 & $2.05\text{mm}$ & $+0.05 \, \text{mm}$ & \textbf{True} \\ 
Diameter Area 3 & $25.06\text{mm}$ & $+0.06 \, \text{mm}$ & \textbf{True} \\ 
Diameter Hole 1 & $14.97\text{mm}$ & $-0.03 \, \text{mm}$ & \textbf{True} \\ 
Height Area 3 & $1.95\text{mm}$ & $-0.05 \, \text{mm}$ & \textbf{True} \\ 
Flatness Surface 1 & $0.10\text{mm}$ & $\pm 0.10 \, \text{mm}$ & \textbf{True} \\ \hline
\end{tabular}
}
\caption{Selection of measurement results showing feature sizes, deviations and tolerances}
\label{tab:measurements}
\end{table}

In order to validate the concept and the usage of the developed NFT process, a milled exemplary workpiece was used. 
The aim of this process was the semi-automatic generation of the NFT from the milling process and its real-time simulation. 
For this purpose, features of the standard workpiece \textit{Diamond Circle Square}\footnote{ISO 10791:2020-7} were defined, which determine its production quality, e.g. shape deviations or surface flatness. 
Using an real-time simulation of the manufacturing process, a 3D model of the manufactured workpiece was created. 
Subsequently, a manual comparison was performed between the generated 3D model and the target dimensions and tolerances. This comparison yielded a list containing the deviations (Table~\ref{tab:measurements}). 

\vspace{-0.2cm}
\begin{figure}[!ht]
    \centering
    \includegraphics[width=0.7\linewidth]{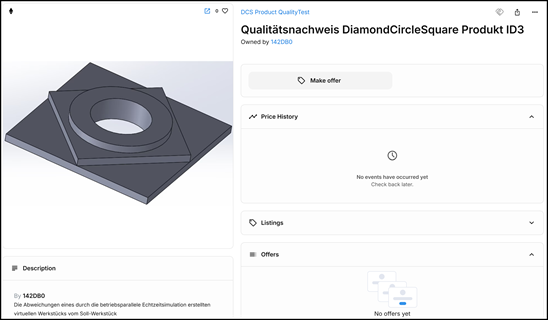}
    \caption{Representation of the minted NFT as proof of quality}
    \label{fig:opensea}
\end{figure}
\vspace{-0.6cm}

Based on this data, an AAS including the \textit{Quality Control for Machining} submodel was automatically created and serialised in the form of a JSON file. 
This file was then encrypted and uploaded to a decentralised IPFS instance\footnote{https://pinata.cloud/}. 
The resulting hash of the file was subsequently utilised as a pointer to the AAS of the minted NFT. 
Having minted the NFT in an Ethereum test environment, it was possible to explore both the NFT and the smart contract on the blockchain. 
Fig.~\ref{fig:opensea} provides a visual representation of an NFT on a commonly used trading platform\footnote{https://opensea.io/}, including the ID of the workpiece, an image, a description, the addresses of the owner and the smart contract collection (\textit{DSC Product QualityTest}) and potential history entries for the Token. 
The page also contains the NFT's publicly available metadata, including the IPFS hash value, which is utilised to locate the actual data of the AAS (Listing \ref{lst:nft_metadata})\footnote{https://ipfs.io/ipfs/Qmb2Bf5Tvh37knbMzsZmWRKpxpmVeA1zkEPNWu3bPMh3YQ}.

\vspace{-0.1cm}
\begin{lstlisting}[language=json, caption={Metadata of a minted NFT}, label={lst:nft_metadata}]
{
  "name": "Proof of quality DiamondCicleSquare product ID 3",
  "description": The deviation of a virtual workpiece created by the operational-parallel real-time simulation from the target workpiece,
  "image": "ipfs://Qmb2Bf5Tvh37knbMzsZmWRKpxpmVeA1zkEPNWu3bPMh3YQ",
  "attributes": [],
  "hidden content": {
    "ipfs_hash": "ipfs_hash://QmUZckYBUheatzsp7E8WQawGmBJdVC77B5rAZbK6EkZejN",
  }
}
\end{lstlisting}

The metadata of the NFT contains the IPFS hash of the encrypted JSON file that is stored publicly. 
When this hash is entered into a browser, an encrypted file is displayed, which is unreadable in this state. 
After decrypting it with the correct symmetric key, the AAS data can be accessed, containing the complete JSON file with the quality-relevant metadata. 
Listing \ref{lst:aasdata} shows quality data for the feature \textit{Height Area~3}. 
It shows a measured value of 1.95 mm, which is within the required tolerance and can be accessed by authorised parties using the key. 

\vspace{-0.1cm}
\begin{lstlisting}[language=json, caption={Partial excerpt of the decrypted JSON file}, label={lst:aasdata}]
{
  "MetrologyData_Height_Surface_3": {
            "QualityActualValue": "1.95",
            "Description": "Total height of surface 3",
            "QualityInSpec": "True",
  }
}
\end{lstlisting}

\section{Conclusion}
The validation demonstrates that the developed concept is capable of automatically generating and modifying an NFT as a proof of quality for existing manufacturing deviations, enabling the trustworthy exchange of data from an online-quality assurance system. 
Since the proposed online-quality assurance is still in development \cite{chen2024utilizing}, the process - from data collection to NFT minting - cannot yet be fully automated, requiring manual processing of manufacturing deviations.
From this point, the minting of the NFT is then fully automated. 

The smart contract developed in this work enables new metadata to be referenced to an existing NFT, allowing multiple companies involved in processing a workpiece to flexibly add their quality-related data. 
The AAS is utilised as a standard information model to ensure interoperability among the saved data. 
Once data is added, it cannot be deleted or modified without being noticed, promoting trustful collaboration as all participants can be confident that the data remains secure from manipulation or deletion by unauthorised parties. 
Overall, this approach minimises the manual effort required to ensure the quality of features for each workpiece or random samples.

The study also found that using a token-based quality management system on a public blockchain is impractical without privacy measures like encryption. 
Given the public accessibility of data on blockchains and IPFS, implementing additional security measures, such as encrypted storage, is essential.
Alternatively, private blockchains could be used to enhance data security and control access.
Also, additional technologies, such as IPFS, need to be integrated to keep transaction costs within a reasonable range. 
Nonetheless, the NFT itself functions as a data verification anchor, irrespective of the referenced data.
This makes it a decentralised, manageable data representation of the product for collaborative quality assessment.

\section*{Acknowledgment}
The authors would like to thank the Federal Ministry for Economic Affairs and Climate Action (BMWK) for funding the joint project growING under reference number 13IPC036G, and the Federal Ministry of Education and Research (BMBF) for funding under reference number COSMIC-X 02J21D140. The presented results were partly created in these projects.

\bibliography{bib}

\end{document}